\documentstyle[epsf,here]{article}
\begin{document}
\baselineskip 1.5\baselineskip

\title{ Detection of coherent vorticity structures using time-scale resolved acoustic spectroscopy}

\author{ Christophe Baudet, Olivier Michel, \\ 
{\small Laboratoire de Physique (URA 1325 CNRS), \'Ecole Normale 
Sup\'{e}rieure de Lyon,}\\
{\small 46 all\'{e}e d'Italie, 69364 Lyon Cedex 07, France}\\
{\small tel~: (33) 472 72 83 78 - fax~: (33) 472 72 80 80}\\
{\small e-mail~: baudet@physique.ens-lyon.fr, omichel@physique.ens-lyon.fr}\\ \\
{ William J. Williams}\\
{\small Department of Electrical Engineering and 
Computer Science,}\\
{\small University of Michigan, Ann Arbor, MI 48109-2122, USA}\\
{\small e-mail~: wjw@gabor.eecs.umich.edu}
}        
\maketitle

\centerline{
Submitted to : {\it Physica D}
}
\vspace{10mm}


\vspace{10mm}

{\bf PACS}


47.32. {\it Vortex dynamics}


47.27.Gs {\it Turbulent Flows - Isotropic turbulence}


43.60. {\it Acoustic signal processing} (43.60.Rw, 43.60.Gk)

\vspace*{20mm}
\rightline{
\underline{{\small E}N{\large S}{\Large L}{\large Y}O{\small N} $\bullet$
{\sc physique}}}

\newpage

\begin{abstract}
It seems widely accepted by the turbulence community that the intermittency 
observed in fully turbulent flows is closely related to the existence of 
intense vorticity events, localized in time and space, also known as coherent structures.
We describe here an experimental technique based on the acoustic scattering 
phenomenon allowing the direct probing of the vorticity field in a turbulent 
flow. In addition, as in any scattering experiment, the information is in the Fourier domain : the
scattered pressure signal is a direct image of the time evolution of a well specified spatial Fourier mode
of the vorticity field. Using time-frequency  distributions, recently introduced in signal  analysis theory,
for the analysis of the scattered acoustic signals, we show  how the legibility of these signals is
significantly improved (time resolved spectroscopy). The method is illustrated on data extracted from a
highly turbulent jet flow :  discrete vorticity events are clearly evidenced. The definition of a generalized 
time-scale correlation function allows the measurement of the spatial correlation length of these events and reveals a 
time continuous transfer from large scale towards smaller scales (turbulent cascade). We claim
that the recourse to time-frequency  distributions leads to an  operational definition of
coherent structures  associated with phase stationarity in the time-frequency plane. 
\end{abstract}

\newpage

Intermittency \cite{Frisch} is usually revealed by analyzing the statistical
properties of the flow deduced from single point measurements 
(e.g. the time record of the flow velocity at one point obtained by hot wire
anemometry). In the past few years, the turbulence community has reached a 
consensus about an acceptable definition of intermittency, this latter being 
defined as deviations (or corrections) from the power law scaling of the 
longitudinal velocity structure functions in the inertial range:
$S_p(l) = \langle (\delta v_{\Vert}(l))^{p} \rangle \propto l^{\zeta_p}$
initially proposed by Kolmogorov in 1941 (K41 theory). Indeed, the K41 theory 
predicts that $\zeta_p = \frac{p}{3}$, whereas clear deviations from this linear law are
experimentally observed \cite{Benzi93}. Recent results suggest that intermittency  is closely
related to the presence of localized (in time and space) vortices in the turbulent flow. 
Such objects, presumably with a filamentary shape, have been observed in some experiments
\cite{DCB91,DPF96} and numerical simulations \cite{Brac91,Les97}. Theoretical models \cite{Lev94}
assuming a hierarchy of coherent structures in the form of vortex filaments, predict a relation 
for the $\zeta_p=f(p)$ in very good agreement with experimental data.
However, a clear definition of coherent vorticity structures is still lacking.\\

There exists a vast amount of theoretical work dealing with the acoustic scattering phenomenon by
velocity fields. Most of it was initiated nearly fifty years ago and started  with the papers
authored by A.M. Obukhov \cite{Obuk53}, R.H. Kraichnan \cite{Kraic53}, Chu and Kovasznay
\cite{Chu58}, to mention but a few. In a more recent paper, F. Lund \cite{Lund} has established,
under reasonable assumptions, a linear relation between the scattered amplitude of a plane acoustic
wave incident on a turbulent flow and the spatial Fourier transform of the vorticity field.
The purpose of this paper is to demonstrate how one can take advantage of the spectral nature 
of the measurement process, in an acoustic scattering experiment, for the detection of 
spatially and temporally localized vorticity events. Indeed, there are numerous examples in 
physics of disordered media, where scattering techniques (e.g. light and neutron scattering in
condensed matter physics) provide useful and pertinent informations about the statistical
properties of these systems. \\

This experimental work aims at introducing, on experimental basis, a
new method for investigating turbulent flows, serving as an alternate to usual single point
measurements like hot wire anemometry. We place the emphasis on some experimental evidence of the
existence of coherent and localized (i.e. intermittent in both space and time) structures. Using a
dual channel setup (acoustic interferometer), we are able to experimentally demonstrate the
effectiveness of the length scale selection (through the selection of a well defined spatial
scattering wave vector) involved in acoustic scattering measurements. Further, as intermittency
implies non stationarity of physical quantities (think about a vortex as being localized in both
time and frequency domains, appearing or disappearing within the analyzed volume), the second part
of this paper focuses on non stationary spectral analysis. The proposed approach is based 
on the recently introduced \cite{will92} Reduced Interference time-frequency Distribution (RID),
of which the properties are briefly discussed. RID analysis is then applied to signals from an
experiment of acoustic scattering by a fully turbulent jet flow. On a single RID reprentation of the 
scattered pressure signals, one clearly oberves well localized energy packets at a given spatial 
wavevector. In order to get further insight on these objets, we compare the RID representations at two
different wavevectors simultaneously recorded. This dual representation demonstrates a significant
correlation of the time occurences and of the velocities (correlated phase derivatives due to correlated
doppler shifts). As this time-frequency correlation decreases with the separation between the two
wavevectors,  we conclude that the different patterns selected at different wavectors belong to what can be
defined as a coherent structure. Notice here that our approach closely matches the definition of
intermittency formerly proposed by  G.K. Batchelor in 1953 \cite{Batch53}: {\em "The inference, then, is
that there is an  uneven distribution, in space of the energy associated with the large wave-number
components of the turbulence, and that the higher the wave-number, the more does the associated energy tend
to occur in confined regions of space (meaning that if a Fourier resolution of the velocity field {\it
within } a region of activity were made, the amplitude of the component at the relevant wave-number would
be found to be large, while a Fourier resolution of the velocity field {\it within } a region of quiescence
would give a very small amplitude; the amplitude of the component for the field as a whole will lie
somewhere between these two amplitudes)"}.
The paper is organized as follows~: part 1 is devoted to the presentation of some
theoretical aspects and assumptions underlying scattering of acoustic waves by a turbulent flow field with
a particular emphasis on the role played by the vorticity field in this interaction. In part 2, we describe
the experimental setup as well as the statistical properties of the turbulent jet flow under investigation.
The data acquisition procedure and the preprocessing of the recorded data are detailed in part 3. Part 4
deals with some theoretical issues of time frequency analysis and the role of the ambiguity function for
introducing the RID. Finally, experimental results are displayed and discussed in part 5.

\section{Acoustic scattering by a turbulent flow}
%

As quoted by Nelkin \cite{Nelk94}, vorticity is intrinsically a non-local quantity, and therefore it
can hardly be measured by using local probes (e.g. hot wire anemometers) \cite{Wallace}.
Although vorticity could "conceptually" be measured by local probes, the main problem resides in
that one needs at least two probes, separated by a finite distance in order to compute the gradients
involved by the curl operator. From an experimental point of view the situation is even worse: at
least nine probes are needed to reconstruct the three components of the vorticity, within the
appropriate Taylor Hypothesis (see \cite{Wallace}). Since these probes have a finite extension and 
finite time response, the computation of the gradients at the smallest scale of the turbulent flow 
becomes problematic \cite{Tsin92,Anton97}.  It is worth noticing here that an alternate experimental has
been proposed by Noullez \& al. \cite{Noullez97} in order to probe directly the spatial transverse velocity
increments in a turbulent flow.
\\

Taking these latter remarks into account, one can find several theoretical \cite{Obuk53,Kraic53,Chu58}, experimental \cite{Engler,Kov76,Korm80,Baerg65,Baud91,BH96} and numerical works \cite{Colon} addressing the direct measurement of the vorticity field using acoustic scattering. As in any
scattering experiment, one expects to be sensitive to the spatial Fourier distribution of the scatterers (as the
amplitude of the scattered  wave is expressed  in terms of some coherent average on ``secondary sources'') over a
finite volume  defined as being the cross section of the acoustic beams (incident and scattered). In a recent paper,
Lund {\it et al.} \cite{Lund} derive a linear relation between the Fourier transform of the scattered ultrasound
amplitude and the  space-time Fourier transform of the vorticity field. In analogy with the more usual light scattering 
phenomenon, the physical mechanism at the origin of acoustic scattering by vorticity can be thought 
of as follows~: an acoustic wave impinging on a vorticity distribution induces fluctuations of the 
vorticity at the incoming sound frequency (by virtue of the Kelvin circulation theorem). Each scatterer (vortex) 
acting as a secondary source will, in turn, radiate a sound wave. The coherent average (taking into account the
relative positions of the individual vorticity elements) over the scatterers distribution results in the emission,
outside the vorticity domain, of a scattered acoustic wave.
It is important to emphasize here that many vortices interact simultaneously with the wave front. Such interactions
have been numerically investigated in \cite{Colon} in the case where a unique vortex is present. In a scattering
process, one is sensitive to spatial details of the scatterers distribution at a typical length scale measured by
the wavelength. Note that whereas a light scattering process is usually linear, the acoustic scattering
phenomenon depicted here stems from the non-linear term of the Navier-Stokes equation, and requires much explanations
and details computations that are beyond the scope of this paper. In particular, one can find in \cite{Lund} a clear
explanation of the respective contributions to the acoustic scattering cross-section of the vorticity field on one
hand, and of the irrotational velocity field, induced by the vorticity, on the other. \\
 
Using a Born approximation, Lund {\it et al.} obtain the following linear relation between the scattered acoustic
pressure amplitude and the spatial Fourier transform of the vorticity field:
\begin{equation}
\frac{p_{scat}(\nu)}{p_{inc}} = \pi^{2} i \frac{-cos(\theta_{s})}{1-cos(\theta_{s})}
\frac{\nu e^{i\nu D/c}}{c^{2}D} (\vec{n} \land \vec{r}).
\vec{\Omega}(\vec{q}_{scat}, \nu - \nu_{o})
\label{eqn_Diff}
\end{equation}
where $\land$ and ``.'' stand for the vector product and the scalar product respectively, and
\begin{equation}
\vec{q}_{scat} = \frac{2\pi}{c}(\nu \vec{r} - \nu_o \vec{n}) \simeq 
\frac{4\pi\nu_o}{c}\sin(\frac{\theta_{s}}{2}) \frac{\vec{r}-\vec{n}}{|\vec{r}-\vec{n}|}
\mbox{ for } \nu \simeq \nu_o
\label{q_scat}
\end{equation}
In these formula, 
\begin{itemize}
\item $\vec{q}_{scat}$ is  the scattering wave vector
\item $\vec{n}$ and $\vec{r}$ are the
impinging and  scattered wave directions, respectively
\item $\nu_o$ is the frequency of the incoming sound 
\item $\nu$ is the frequency of
the scattered sound
\item $\theta_{s}$ is the scattering angle 
\item $p_{inc}$ stands for the pressure amplitude of the
incoming sound wave (assumed to be plane and monochromatic)
\item $p_{scat}$  is the amplitude of the scattered sound wave
\item $D$ and $c$ stand for the acoustical path between the measurement area and the detector, and the
sound velocity respectively.
\item $\vec{n}$ and $\vec{r}$ are the unit vectors in the direction of the probing and detected sound waves
\end{itemize}
It is important to emphasize here that among the hypothesis used in \cite{Lund} the  characteristic 
time of any process within the flow has to be very large compared with the period of the analyzing sound wave. 
This latter remark implies that $\nu \simeq \nu_o$ and that this method is  relevant only for flows with low Mach
number. Equation (\ref{eqn_Diff}) shows that, using acoustic scattering, it is possible to probe one component
(perpendicular to the scattering plane defined by the vectors $\vec{n}$ and  $\vec{r}$) at a chosen length scale
through the selection of a known scattering wave vector $\vec{q}_{scat}$.\\

In a previous similar experiment \cite{Baud91} (see also \cite{Gromov82}) performed on the regular vortex street
behind a cylinder at low Reynolds numbers, we have demonstrated that the scattering cross section is indeed
strongly dependent on the direction of the vorticity field accordingly to Equation (1) of the present paper.
This latter experiment evidences the existence of  direct interactions between vorticity and
acoustic waves~: among the nine components of the stress tensor only one vector can be defined
without ambiguity, i.e. the vorticity (antisymmetric part of the tensor). Furthermore, a nice
feature of this method lies in the fact that it is a non perturbating technique.

\section{Experimental setup}

The acoustic waves are produced and detected by Sell-type transducers (see e.g. \cite{Anke}),
the dimensions of which are 15cm x 15cm. The advantage of dealing with such large transducers
is to minimize diffraction effects and therefore to enable a sharp selectivity of the  scattering
wave vectors ($\vec{q}_{scat}$ defined above), both in modulus and direction \cite{Baud91}.
As shown on figure (\ref{setup}), the direction of $\vec{q}_{scat}$ is aligned 
with the mean-flow velocity~ in order to maximize the Doppler shift. The magnitude of $\vec{q}_{scat}$, 
and thus the probed length-scale in the flow, is then selected   by tuning the incoming sound 
frequency $\nu_o$, the scattering angle being held fixed.
Indeed, for practical purposes, it is much simpler to hold the geometry of the experimental setup
constant.  Therefore, $\theta_{s}$ is constant and $\vec{q}_{scat}$ has constant direction. The 
scattering angle $\theta_{s}$ was set to 60 degrees, as a trade-off between optimizing the
sensitivity of the measurements and limiting the spurious effects induced by diffraction side-lobes 
\cite{Baud91}.
It can then be seen from (\ref{eqn_Diff}), that the Fourier transform (FT) of the scattered pressure 
signal is proportional to the band-pass  spatial frequency filtered vorticity field. The central frequency 
of the latter filter is $q_{scat}(\nu_o)=||\vec{q}_{scat}(\nu_o)||$. Its width ($\Delta q$) can be  obtained from the diffraction 
pattern of the Sell transducers \cite{Baud91}. A rough estimate of $\Delta q$ is given by differentiating
$\vec{q}_{scat}$ with respect to $\theta_{s}$, which leads to 
$\Delta q = \frac{2\pi\nu_o}{c}\cos(\theta_{s}/2).2\Delta\theta$, where $\Delta \theta$ is given by 
a diffraction ansatz: $\Delta\theta \simeq \frac{c}{\nu_o L}$, $L$ being the characteristic size of the transducer 
\cite{KFCS82}. The volume  which is analyzed is defined by the cross section of the incident beam and the
main lobe of the receiver (antenna beam). Its dimensions in our experimental setup were of the same
order of magnitude (a bit larger) than the transducers dimensions.\\
 The flow which we investigate in this paper, consists in an axi-symmetric turbulent jet flow ($R_{\lambda}
\simeq 600$).  emerging from a circular nozzle of diameter 5 cm, at a high velocity (about 60 ms$^{-1}$).
Details about our wind tunnel facility can be found elsewhere \cite{Benzi93}. The volume of interest was
set at a distance of  about 50 nozzle diameters downstream ensuring that the turbulence is fully developed.
A study based on classical anemometry (performed in the same area of interest)  leads to the estimated
following parameters~: the average longitudinal velocity is about 5.3 ms$^{-1}$ and the turbulence ratio is
about 28\% \cite{BH96}. The third order structure  function exhibits a clear power law scaling over 1.5
decades (from the Taylor scale $\lambda = 6$ mm up to the integral length scale $l_o = .25$ m), with an
exponent close to 1 in accordance to the Kolmogorov four-fifths law
\cite{Frisch}. From this inertial range scaling, we deduce the value of the mean energy dissipation per unit mass
$\epsilon = 13$ m$^2$s$^{-3}$, the Kolmogorov micro scale $\eta = 131$ $\mu$m and the large eddy turnover time $t_o =
168$ ms. The analyzing  sound frequency can be varied from 5kHz up to 200kHz (where sound attenuation effects cannot
be neglected). Thus we are able to probe a large part of the inertial range, as well as length scales belonging to the
beginning of the dissipation range \cite{BH96}. However in this paper, we shall restrict our analysis to a single 
length scale pertaining to the inertial range ($\frac{q_{scat}}{2\pi} = 0.6 $ cm$^{-1}$ corresponding to a typical
length scale of 1.7 cm).

\section{Data acquisition and pre-processing}
\subsection{Single channel measurements}

An experiment consists in measuring and recording time series of the acoustic pressure signal
received on a Sell-type transducer (see above). The emitted sound wave can be considered as
being a monochromatic plane wave (if diffraction effects are neglected at first sight). 
The sampling device (HP3565) has 16 bit resolution and its sampling frequency is 262144 Hz.
Each recording channel is provided with its own local oscillator (tuned at $\nu_{o}$), thus enabling a
numerical heterodyne  detection of the scattered signal around the frequency of the impinging
sound wave. Note also that, in order to reduce phase noise along the whole measurement line, the waveform
generators (sound emission), the sampling clock and the local oscillators (demodulation) are locked on the same 10
MHz master clock. Two time series associated with synchronous phase and quadrature signals are recorded at a rate of 8192 samples per second, from which the analytical complex signal is computed (characterized by a phase and an  amplitude). Indeed,
thanks to the linearity of our ultrasonic detectors, the phase information of  the scattered pressure wave is preserved
(the phase reference being that of the incoming sound wave). \\
As indicated by equation (\ref{eqn_Diff}) (expressing a convolution
in the time Fourier domain), the scattered  pressure signal is obtained as a time modulation of the incoming pressure
signal (carrier wave) by the low frequency  signal associated with the vorticity of the flow under investigation (for
small enough Mach numbers).  One thus expects a narrow band scattered signal with frequencies close to the incoming
sound frequency $\nu_o$ (see figure \ref{spec_fig} a). Then, the purpose of the above mentioned heterodyne detection
is  to significantly reduce the amount of stored data.
Equations (\ref{eqn_Diff},\ref{q_scat}) lead (for $\nu \simeq \nu_o$) to express the 
Fourier transform of the detected pressure signal as
\begin{equation}
p_{scat}(\nu) \propto\frac{\cos(\theta_s)\cos(\frac{\theta_s}{2})}{q_{scat}}\vec{\Omega}_{\perp}(q_{scat},\nu-\nu_{o})
\label{eqn_Diff2}
\end{equation}
where $q_{scat}$ and $\vec{\Omega}_{\perp}$ stand for the magnitude of the scattering wave-vector 
and the vorticity component orthogonal to the scattering plane, respectively. We have represented on 
Figure \ref{spec_fig} (a) a typical averaged power spectral density of the scattered pressure signal 
(after demodulation) obtained in the turbulent jet flow. Notice that in this figure and throughout this paper, 
the frequency axis represents the frequency shift with respect to the incoming sound frequency $\nu_{o}$.
The spectrum is clearly non-symmetric with respect to the null frequency. This latter fact is a simple
consequence of the advection of the vorticity field by the flow velocity and to our decision to align the
scattering wave vector with the direction of the mean flow velocity $\vec{V}_{mean}$. Indeed, the
scattering of waves by a moving structure gives rise to a frequency shift
$\Delta\nu$ as a consequence of a Doppler effect \cite{Morse} :
\begin{equation}
\Delta\nu=\nu-\nu_{o}=\frac{1}{2\pi}\vec{q}_{scat}.\vec{V}_{mean}
\label{Doppler}
\end{equation}

Notice that this latter relation can be obtained by applying a Galilean transform in the real space 
followed by a double Fourier transform on the space and time variables.\\

From (\ref{eqn_Diff2}) one expects a divergence of the scattered signal around $\theta_s = 0$, 
for which $q_{scat}=0$ and thus $\Delta\nu=0$. Indeed, the existence of direct paths between the emitter
and receiver (associated to diffraction effects due to the finite size of the transducers) results in the
presence of a very strong peak in the spectrum $p_{scat}(\Delta\nu)$ around $\Delta\nu=0$.
From (\ref{Doppler}), we are able to relate $q_{scat}$ to $\Delta\nu$, and thus get the following 
expression for the received acoustic signal
\begin{equation}
p_{scat}(\nu_0 + \Delta\nu) \propto \frac{1}{\Delta\nu}\vec{\Omega}_{\perp}(q_{scat}, \Delta\nu)
\label{eqn_Diff3}
\end{equation}
Bearing in mind that the scattered pressure signal is recorded after a numerical 
heterodyning procedure that shifts its analytical frequency spectrum around the null
frequency (the frequency of local oscillator in the lock-in is set to $\nu_{o}$), one gets
\begin{equation}
P_{scat_{rec}}(\Delta\nu) \propto \frac{1}{\Delta\nu}\vec{\Omega}_{\perp}(q_{scat}, \Delta\nu)
\label{eqn_Diff4}
\end{equation}
where $P_{scat_{rec}}$ stand for the scattered pressure recorded signal. 
Taking the inverse FT of equation (\ref{eqn_Diff4}) leads to 
\begin{equation}
P_{scat_{rec}}(t) = \int_{-\infty}^{+\infty} \vec{\Omega}_{\perp}(q_{scat}, t)dt
\end{equation}
The preceding equation expresses the recorded scattered signal (after heterodyning) as being the integral of 
$\vec{\Omega}_{\perp}(q_{scat},t)$ with respect to the time variable 
As a consequence, the recorded data must be differentiated to give a correct representation of
$\vec{\Omega}_{\perp}(q_{scat},t)$. It is important to notice here that the latter differentiation must 
be performed at a low frequency, since the  heterodyne detection has already been applied. 
The averaged power density spectrum (PDS) of $p_{scat}(t)$ recorded for the jet flow (see preceding section) 
and $\nu_{o}=32kHz$, is shown on figure  (\ref{spec_fig} a). 
Figure (\ref{spec_fig} b) illustrates the efficiency of the difference  filter applied to the time series 
for extracting a PDS proportional to $\vec{\Omega}_{\perp}(q_{scat},\Delta\nu)$.
It is worth noticing at this point that this differentiation ansatz is nothing else than
a transposition of the Taylor hypothesis (heavily used in single point
measurement). Actually, it suffers from the same limitations with regards to the level of turbulence.However, as 
explained below, these limitations can be overcome by the resort to the  time-frequency distributions.

\subsection{Two-channels measurements} 
According to (\ref{q_scat}), a given spatial spectral component of the vorticity 
field may be probed by using different sets $(\theta_s, \nu_0)$, associated with a couple 
of transducers (one emitter, one receiver) which will be referred to as a {\em channel} in the rest
of the paper.
The acoustical paths corresponding to the null scattering angle for which  divergence occurs 
(thus explaining the presence of a strong peak at zero frequency shift, see (\ref{Doppler})) is 
then different for each set of  ultrasonic emitter-receiver. As a consequence, it is expected that the parts of the
signals  corresponding to the null scattering angle, recorded on each channel are mutually non coherent.
Following the same reasoning, it is expected to find a high degree of coherence 
between parts of signals that are due to scattering effects from the same measurement area (cf 
preceding sections), provided that both channels are tuned to probe the same scattering wave-vector:
\begin{equation}
\frac{4\pi\nu_{0,1}}{c}\sin(\frac{\theta_{s,1}}{2})=\frac{4\pi\nu_{0,2}}{c}\sin(\frac{\theta_{s,2}}{2})
\label{tuning}
\end{equation}
where $\nu_{0,i}, \theta_{s,i}$ stands for the ultrasound frequency and scattering angle associated with  channel $i$.

The additional hypothesis in the 2-channel measurement scheme, is that the measurement areas defined by the 2 sets of
transducers are actually identical. This point is crucial in our experiment, as any mismatch in the measurement areas
would preclude any possibility to compare scattering signals, which would then be associated to different spatial
distributions. The setting of the experiment is derived by carefully aligning the transducers with laser diode rays.
Then, the lengths of the acoustical paths for each channels are  set to the same value (up to a precision of
1mm)\footnote{ The propagation time of acoustic waves over this distance is about $3.10^{-6} s$, i.e. much
smaller that the sampling period.},
in order to insure equivalent propagations delays on both channels. This is achieved by checking the phase of the cross
channels inter-spectrum estimates, and shifting the transducers along its  main diffraction axis accordingly: for null 
propagation delay, the slope of the phase with respect to frequency must be zero, as group delays are expected to be
identical on both channels. \\
Figure (\ref{figCoh}) illustrates the fact that the divergences at zero-frequency shifts, 
measured on two channels verifying (\ref{tuning})  are mutually non coherent, whereas the part 
of the spectrum corresponding to the scattered waves from the same  measurement 
area (Doppler shifted)  are highly similar (high cross-coherence values). Furthermore, the values obtained 
for the cross-coherence do not depend on whether the differentiation (see preceding paragraph) was performed,
as the same linear filter (time differentiation) is applied to  both channels. 
This latter observation illustrates the pertinence of the differentiation for extracting the interesting part 
(i.e. the signature of the vorticity) in the recorded signal. The phase of the cross-channels inter-spectrum is shown on
the last plot of figure (\ref{figCoh}): one sees that  the phase slope takes low values within the frequency domain
where the spectrum of  scattered signals is significant.\\
So far, we have dealt with second order statistical properties of the turbulent flow, and considered 
long enough records, to insure that the stationary hypothesis was correctly met. However, the spectrum of the recorded
time-series corresponding to one spatial  frequency component of the vorticity field, turns out to have an important
width (Equivalent 10dB bandwidth= 400Hz).
This may be ascribed to either fluctuations of the advection velocity (thus leading to fluctuations of the 
Doppler shift), or to finite duration of vorticity structures. In this latter case, one then expects to record
"sine-wave trains' of finite duration. Notice that both effects are likely to occur simultaneously.
Therefore, one is naturally led to look for an analyzing tool allowing a joint representation in 
time (in order to access ``vortex life time'') and frequency (for evidencing Doppler shifts, 
or equivalently, advection velocity fluctuations). 

\section{Time-frequency analysis}

When finite duration events of vorticity are present, the scattered acoustic 
signal will exhibit a time-varying spectrum, as the frequency of the measured pressure
signal drops from $\nu_0$ to $\nu_0 + \delta\nu$ when a vorticity event exists, 
that has energy in the spatial Fourier domain at the wave-vector $\vec{q}_{scat}$.
Due to the randomness of the velocity field which advects the vorticity field, the finer 
details of the time dynamics of individual events are wiped out when average spectra of 
the scattered signals are computed. Indeed, on the average, the scattered spectrum has a Gaussian shape
reflecting the statistical properties of the flow velocity \cite{BH96}. 
Notice that an infinitely thin and localized vortex (theoretical) would have contributions
at all wave-vectors in the spatial Fourier domain.

Our aim is thus twofold~: first, detecting presence/absence of such events in the recorded signal,
and second, estimating the frequency shift $\Delta\nu$ at which this occurs, in order to characterize
its relative velocity. We are thus faced with the need for a representation of the signal which preserve
simultaneously the time and frequency information.\\ 
Temporal fluctuations of the spectral distribution can be approached
by  the common spectrogram analysis. Watkins \cite{wat66}, among others, 
has addressed this problem and provides examples of reasonable and unreasonable spectrogram results
in the analysis of bio-acoustics signals (marine mammal sounds). However, the spectrogram involves a
sliding time window which intends to capture a portion of  the signal which is sufficiently
restricted in time so that stationarity assumptions  are approximately met. Furthermore, the
presence of the window results in a time-frequency distribution (TFD)
exhibiting both temporal and spectral leakage. \\
The Wigner-Ville distribution (WVD) avoids the problems of windowing the signal
but the presence of interference terms between signal components, due to the quadratic nature
of the WVD (see next paragraph)  \cite{fla84,hla84} often precludes its applicability. 
Recent studies have shown that a class of
smoothed WVD, the Cohen's Class of TFDs, allows to reduce the amplitude of the
interference terms, while preserving covariance properties of the WVD in both the 
time and frequency domains. Cohen \cite{coh89,cohen95} and Flandrin \cite{Fla93} have
recently  provided  excellent reviews of TF analysis, including recent developments.
It is worth noticing here that the above mentioned signal representations do not rule out
the Gabor-Heinsenberg inequalities (see \cite{fla84}).
A brief presentation of the main issues on 
Cohen's class TF analysis and the key role played by the ambiguity function is given below. \\
In this paper, we focus on one distribution from the Cohen's class, the Reduced-Interference Distribution
(RID). It was first introduced by Choi and Williams \cite{choi89}, and has the additional feature of
preserving the time and frequency support of signal components~: ($RID(t,f)$ is non zero if
the signal is non zero at $t$ and has a spectral component at $f$). These are highly
desirable properties if one wishes to infer statistical properties of the vorticity distributions
from this TF analysis. The examples shown were computed with a binomial RID, which is a good
discrete approximation of the continuous distribution \cite{Jeong92}.

\subsection{The WVD and Ambiguity function}
The Wigner-Ville distribution of $z(t)$ is defined as the Fourier transform (F) of 
$$R_z(t,\tau)~=~z(t+\frac{\tau}{2})z^{*}(t-\frac{\tau}{2})$$
with respect to the lag variable $\tau$. 
\begin{equation}
W_z(t,\nu)=F_{\tau}[z(t+\frac{\tau}{2})z^{*}(t-\frac{\tau}{2})]=F_{\tau}[R_z(t,\tau)]
\label{e1}
\end{equation}
where $z(t)$ is the time signal, $z^{*}$ is its complex conjugate. The operator
$F_{(.)}$ denotes the Fourier transform operator, with respect to the variable $(.)$;
this notation will be adopted in the rest of that paper. From an experimental and heuristic 
point of view, the effect of the operator $R_z(t,\tau)$ is equivalent to a local phase conjugation of 
the signal which enhances the phase derivative related to the Doppler shift (P. Flandrin : private communication). 
Similarly, but a with different physical meaning, the symmetrical\footnote{
Ordinarily, the AF is defined as the FT of $R_z(t,\tau)$, with respect to $t$. The symmetric
AF used here is defined as the inverse FT of $R_z(t,\tau)$, with respect to $t$. This does not
induce any change in the interpretation of the AF (see \cite{cohen95}), and is only a matter of convention.
 However, it allows to  relate the AF 
to the WVD via a two-dimensional FT with respect to $t$ and $\tau$ (see equation \ref{WVD-AF}).
}
ambiguity function (AF) is defined as the inverse Fourier transform ($F^{-1}$) of 
$R_z(t,\tau)$ with respect to the first variable\footnote{
 The ambiguity function reduces to a deterministic correlation function in time if 
 $\theta$ is set to zero. Similarly, as can be seen in a dual form which 
 starts with $Z(\nu)$, it  can be
 seen as a deterministic correlation of spectra if $\tau$ is set to zero.
}
\begin{equation}
A_z(\theta,\tau)=F^{-1}_t[z(t+\frac{\tau}{2})z^{*}(t-\frac{\tau}{2})]=F_{t}^{-1}[R_z(t,\tau)]
\end{equation}
Thus, $W_z(t,\nu)$ and $A_z(\theta,\tau)$ are related by the two-dimensional Fourier transform
\begin{equation}
W_z(t,\nu)=\int \int A_z(\theta,\tau) e^{-j(t\theta + \nu \tau)}d\theta d\tau
\label{WVD-AF}
\end{equation}

\subsection{Cohen's Class of Distributions}
Cohen's Class of distributions is defined as 
\begin{equation}
C_z(t,\nu, \phi)=
\int \int \int e^{j((\xi-t)\theta- \nu \tau)}\phi(\theta,\tau)z(\xi+\frac{\tau}{2})z^{*}(\xi-\frac{\tau}{2})d\xi d\tau d\theta
\end{equation}
where $\phi(\theta,\tau)$ is the {\em kernel} of the distribution\footnote{
The range of integrals is from $-\infty$ to $\infty$ throughout this paper.
}.

These relationship can be combined with Eq. (\ref{e1}) to show that 
$C_z(t,\nu,\phi)$ may be expressed as:
\begin{equation}
C_z(t,\nu, \phi)=\int \int \phi(\theta,\tau)A_z(\theta,\tau)e^{-j(t\theta + \nu \tau)}d\theta d\tau
\end{equation}
Thus, while $W_z(t,\nu)$ is obtained from the symmetric ambiguity function by means of 
two-dimensional Fourier transform, any member of Cohen's Class of distributions may be found 
by first multiplying its kernel $\phi(\theta,\tau)$, by the symmetric ambiguity function and 
then carrying out the two-dimensional Fourier transform. 
The spectrogram and the WVD can easily be shown to be members of Cohen's class by choice of 
the proper kernel in each case.\\
The generalized ambiguity function 
$\phi(\theta,\tau)A_z(\theta,\tau)$ is a key-concept in t-f analysis which aids one in 
clearly seeing the effect of the kernel in determining $C_z(t,\nu,\phi)$. 
In the ambiguity domain, application of the kernel of Cohen's class involves multiplication of 
the symmetric ambiguity function by the kernel. This is roughly analogous to application of 
filters by multiplication of the Fourier transform of the signal by the transfer function in 
the frequency domain.
This operation manifests itself as a convolution in the time domain, thus, 

\begin{equation}
 C_z(t,\nu,\phi)=F_{\tau}[\psi(t,\tau)*_tR(t,\tau)]
\end{equation}
where $\psi(t,\tau)=F_{\theta}^{-1}[\phi(\theta,\tau)]$.
Similarly, application of the kernel in the TF domain involves convolution in time and
frequency in  Cohen's class of distributions. Conventional filtering concepts can be
adapted to  Reduced Interference Distribution (RID) design.

\subsection{The RID approach}
New distributions with reduced interference (RIDs)
\cite{choi89,will92,Jeong92} have been developed  in recent years, serving as
attractive alternates for the spectrogram and the WVD.  The RID represents an attempt to
preserve most of the desirable features of the WVD while  alleviating most of the
confusing interference between signal components.\\

RID kernels exhibit low pass characteristics in both the time-shift ($\tau$) and frequency shift 
($\theta$) dimensions in the ambiguity plane, away from the axes. It has been shown
\cite{will92,Jeong92} that any real valued  symmetric, unit area function $h(t)$ which is
limited  to insure time and frequency 
support properties, such that $h(t)=0$ for $|t|>\frac{1}{2}$ and  which has a 
$H(\nu)~|_{\nu=0}=1$ is a RID kernel. One simply replaces $H(\nu)$ with $H(\theta\tau)$ 
in order to have a RID kernel in the ambiguity domain form. Usually interference terms
are located far from the 
$\theta-\tau$ origin, so the filter
can often be designed to provide strong attenuation of these terms, while preserving the desirable 
auto-terms of the individual signal components. The form of the RID kernel for the
$(t,\tau)$ domain application is $\psi(t,\tau)=\frac{1}{|\tau|}h(\frac{t}{\tau})$ and
computation is usually most convenient in this domain.
Preservation of the time and frequency marginals is one of 
the important properties of RIDs. One can recover the
energy spectrum (by integrating with respect to time)  or the instantaneous power (by integrating with
respect to frequency). This is not the case for spectrograms  and many other popular TF distributions
(see e.g. \cite{hla92}).

\section{Application of TF distribution analysis to turbulence signals}
 
A nice discrete kernel which has a binomial form has been used in our applications.
This kernel is computationally efficient and also exhibits desirable characteristics in terms of 
TF distributions \cite{will92}.  The time frequency distribution obtained by
applying the binomial kernel takes the following discrete form:
\begin{eqnarray}
\lefteqn{ {\hat{TF}}_{rid,bn}(n,\nu) = } \\
& & \sum_{\tau=-\infty}^{\tau=\infty} h(\tau) \sum_{v=-|\tau|}^{v=+|\tau|} 
\frac{g(\nu)}{2^{2|\tau|}} \left( \begin{array}{c} 2|\tau| \\ |\tau|+v \end{array}
			\right) z(n+v+\tau)z^{*}(n+v-\tau)e^{-j4\pi\nu\tau}  \nonumber
\end{eqnarray}
where $z[n], n=1 \ldots N$ depicts the sampled time-series under analysis. 
Functions $h(\tau)$ and $g(\nu)$ 
are frequency smoothing window and time smoothing window respectively. $h(0)$ and
$G(0)=TF_v[g(v)]$ are forced to one. Note that these windows were set to
$h(\tau)=\mbox{rect(N)}$  and $g(\nu)=\delta(\nu)$ in the original paper by Williams and
Jeong,  where $\mbox{rect}(N)$ stands for the $N$-points rectangular window, and
$\delta$ stands for the discrete Dirac distribution. The binomial TFD is a good discrete time
approximation of the continuous RID.
The binomial TFD (hereafter called the tf-rid) was computed for acoustic signals obtained in 
the turbulence experiment depicted in the previous sections.  \\
In the cases which are presented, two channels were used to probe the same measurement area~:
the experimental setup is carefully arranged in such a way that the measurement areas defined
by the intersection of the acoustic beams of each channel are actually superimposed. This latter setting
will be maintained for all the experiments presented in the remainder of this paper.
First and second channel had the following characteristics
$(\theta_{s,1}=60 \mbox{ degrees}, \nu_{0,1}=20 kHz)$ and
$(\theta_{s,2}=40 \mbox{ degrees}, \nu_{0,2})$. Applying eq. (\ref{q_scat}) and with the 
transducers arranged as described on figure (\ref{setup}), one shows that the probed scattering
wave vectors are co-axial with the mean flow, and oriented downstream. These latter channel
characteristics correspond to $|q_{scat}|=369.6 \mbox{m}^{-1}$ 
i.e. the corresponding scale (wavelength) is $1.7$~cm, well within the inertial range of the flow.
The same wave vectors are probed onto two different channels by tuning  $\nu_{0,2}$ such
that the condition expressed in eq.(\ref{tuning}) is verified. This  latter condition is met
here for $\nu_{0,2}=29.25kHz$. 
Figure (\ref{fig20294}) shows the tf-rid obtained for the case where the preceding tuning condition
is approximately met (actually, $\nu_{0,2}=29400$ Hz for this experiment). Smoothing windows
$h$ and $g$ were both Hanning windows of respective length $L_h=129$ and $L_g=3$.\\
The frequency marginals, obtained by integrating the tf-rid over time are shown on the left hand
side of the plot. This frequency marginal matches the power density spectrum  (PDS)
(an approximation is introduced by the use of smoothing windows $h$ and $g$). The center of mass
$\Delta \nu_{cm}$ of the PSD is indicated. Replacing $\Delta \nu$ by $\Delta\nu_{cm}$ in 
eq.(\ref{Doppler}) allows one to estimate the average (estimated over 0.6 seconds) axial mean flow 
velocity $|\vec{V}| \simeq 5.57 ms^{-1}$. This result is in good concordance with alternative 
measurements of the velocity (hot wire anemometry). \\
The pertinence of both the tf-rid approach, and the length scale selection, 
within the context of these experiments is illustrated by the high similarity 
between the tf-rid obtained from the two different (and independent) measurement channels. 
Both channels evidence the existence of time localized structures (energy packets in the tf-rid) 
at the same set of instants and with identical Doppler shifts, which was to be expected as the 
probed wave vectors are very close to each other (in modulus and direction) and are defined  
within the same measurement volume (cf figure (\ref{setup})). \\
The interest of such tf-rid representation in the present context is thus threefold. 
First, it emphasizes the fact that the broadening of the averaged spectrum is due to both the short
duration of the structure or events that are the origin of the scattering effects, and the 
fluctuations of the Doppler shifts (the center of mass of the energy packets in the tf-rid
are not well aligned along an axis $\Delta\nu = constant$. 
Second, such representation provides 
a straightforward way to detect the time of occurrence of events at a given scale. A systematic 
study of the statistical distributions of such events, as functions of the probed scale is
presently a matter of great interest and will be presented in a future paper. 
Third, the use of
two simultaneous channel measurements allows one to test the existence of events at different
wavelengths arbitrarily close to each other (scale resolution). In the preceding example, 
the characteristics were set to probe two close scales (namely $|q_{scat,1}|=369,6 m^{-1}$ and $|q_{scat,2}|=371,6 m^{-1}$). 
Structures that are evidenced on one channel clearly appear jointly in the second channel; this latter point 
is illustrated on the central plot of figure (\ref{fig20294}), where the normalized time marginal of the 
geometric mean of the rid of the channels, $d(t)$ is shown:
$$
d(t)=\frac{ \int_{f_n=0}^{f_n=.5} tfr_{rid,1}(t,f_n).tfr_{rid,2}(t,f_n) df_n}{\sqrt{
\int_{f_n=0}^{f_n=.5} tfr_{rid,1}(t,f_n)}\sqrt{\int_{f_n=0}^{f_n=.5} tfr_{rid,2}(t,f_n)}}
$$
$tfr_{rid,i}$ is the tf-rid of the signal recorded on channel $i$, and $f_n$ is the normalized frequency. Note that $0 \leq d(t) \leq 1$ by virtue of Cauchy-Schwarz inequality.
This method (maximum likelihood criterion) allows  detection of the joint presence 
of energy at a given time an d with a given Doppler shift on both channels. Figure (\ref{fig20294zoom}) allows to evidence the similarity of the signals recorded on the channels, even when considered in details, thus revealing similar detection of the local dynamics.\\

The same experiment was performed by resetting the analyzing frequency of the second channel
to $\nu_{0,2}=35 kHz$. Thus, $|\vec{q}_{scat,2}|=442,4 m^{-1}$ with the same geometry, i.e. 
$\vec{q}_{scat,1}$  and $\vec{q}_{scat,2}$ were still colinear, 
aligned with the flow and oriented downstream.
The estimated tf-rids obtained for this situation are shown on figure (\ref{fig20350}). The
normalized co-occurrence function $d(t)$ takes lower values than those obtained in the preceding 
case. Furthermore, the number of co-occurrences that are detected is much lower in this case, 
compared to the co-occurrence detected when the tuning condition (cf eq. \ref{tuning}) is verified.
Further insight is gained by repeating the previous experiments with $|q_{scat,1}|$ held fixed, 
while $|q_{scat,2}|$ is varied according to $\nu_{0,2}$ ranging from $25kHz$ to $35kHz$ (i.e. $ 316m^{-1} \leq |q_{scat,2}|
\leq 442m^{-1}$). For each couple $(\vec{q}_{scat,1},\vec{q}_{scat,2})$, defining a spatial wavector separation 
$\delta q = |\vec{q}_{scat,2} - \vec{q}_{scat,1}|$, we compute the RID representations. Then we introduce a time-scale
generalized cross-correlation function $C(\tau,\delta q)$ defined as:
$$
C(\tau,\delta q)=\left\langle \frac{ \int_{f_n=0}^{f_n=.5} tfr_{\vec{q}}(t,f_n).tfr_{\vec{q}+\vec{\delta q}}(t+\tau,f_n)
df_n}{\sqrt{\int_{f_n=0}^{f_n=.5} tfr_{\vec{q}}(t,f_n)}\sqrt{\int_{f_n=0}^{f_n=.5} tfr_{\vec{q}+\vec{\delta
q}}(t+\tau,f_n)}}\right\rangle_t
$$
where $\left\langle \bullet \right\rangle_t$ stands for an average over the time variable $t$.
Figure (\ref{fig_Corr}.a) sketched the results of this series of experiments on a 2D-plot 
where the amplitude of $C(\tau,\delta q)$ is coded in gray levels, as a function of $\delta q$, and the delay time
variable $\tau$.  The evolution of $C(\tau,\delta q)$ for 3 values of $\delta q$  (namely $\delta q = 0.0, 0.2, -0.2$) is
plotted on Figure (\ref{fig_Corr}.b). Figure (\ref{fig_Corr}.b and .c)  demonstrates that the mean delay time
$\tau$ evolves monotonously from negative values when $\delta q \leq 0$ towards positive values when $\delta q \geq 0$. 
The evolution of the maximum of the correlation function is plotted on figure (\ref{fig_Corr}.d) as a function of
$\delta q$. As expected, the correlation function is maximum for $\delta q = 0$ decreases
smoothly as $|\delta q|$ increases. The observed evolutions of both the value and the sign of the delay time is an
experimental evidence of a time-continuous turbulent cascade of energy from large scales ($\delta q \leq
0$) towards smaller scales ($\delta q \geq 0$). Up to our knowledge, this time continuity of the cascade which
is a direct evidence of the time reversal symmetry breaking had not been yet experimentally demonstrated.
From Figure (\ref{fig_Corr}.b) we measure a typical extension of the detected wave-packets in the q-domain : $\delta q
\simeq 0.4 cm^{-1}$, corresponding to a spatial correlation length of about 10 cm. It is worth noticing here that the
measured correlation length remains smaller than the size of our transducers. It is interesting at this point to mention 
a few words about the spatial and spectral resolutions of the experimental setup. The spatial resolution is determined by
the maximum frequency one is able to propagate through the flow of interest; actually, this frequency is around $200kHz$
enabling to probe lengthscales as small as $1.5mm$ \cite{BH96} (this upper frequency limit is due to sound absorption
phenomena). The spectral resolution, determining the minimum spatial wavevector separation one can probe, depends directly
on the size $L$ of the transducers which fixed the width $\Delta q$ of the equivalent spatial filter \cite{Baud91} ($\Delta
q \simeq \frac{1}{L}$ see also section 2). Notice however, that the latter limitation (analogous to the Rayleigh criterion
used in optics) can be overcome by resorting to a deconvolution scheme once the diffraction pattern of each transducer has
been modelized or characterized.

\section{Conclusion}

We propose a method for evidencing coherent vorticity structures in turbulent flows. 
Though this existence seems to be widely accepted and evidenced by numerical simulations  
\cite{Brac91} or in some peculiar flows \cite{DCB91, DPF96} it still deserves further experimental 
efforts. Furthermore, no clear definition seems to exist for coherent structures.
The experimental technique described here is derived in the spatial Fourier domain and recovers 
the intuitive notion of phase stationarity thus allowing one to understand these 
structures as propagating energy (more precisely enstrophy) packets.\\
The main advantage of the acoustic technique relies on its global and non pertubative character, 
as it gives access to the instantaneous value of spatial Fourier components of the vorticity field.\\
Using two distinct channels (acoustic interferometry) we are able to establish the robustness (noise
immunity) of our vorticity detection scheme. In addition, this configuration enables the simultaneous
probing of two spatial wave vectors arbitrarily close to each other (scale resolved spectroscopy). This latter remark is in 
contrast with usual one-point measurement (hot wire anemometry) in turbulence which do not allow a clear separation of time
and space dynamics. Indeed in such measurements, one has to use the Taylor hypothesis in order to recast time increments
into spatial increments.\\
The joint resort to acoustic scattering and to time-frequency distributions  leads to a measurement
process whereby time (date of occurrence) and frequency (Doppler shifts)  informations are preserved (time resolved
spectroscopy). It is worth noticing here that this kind of analysis could be easily transposed to data obtained from
numerical simulations. Indeed, in the case of pseudo-spectral direct numerical simulations (DNS), a great part of the
computing time is spent in  the spatial Fourier space. 
We claim that time-frequency distributions offer an efficient tool for the detection of coherent structures as it is a
``blind'' detection : no a priori information is needed (for example on the shape of the structures) and one only resorts
to a stationary phase criterion (stating that components of a coherent structure are convected with the same velocity). 
From the RID representations of the scattered signals (simultaneously acquired), we define a generalized time-scale
correlation function based upon a maximum likehood criterion. This time-scale correlation function reveals salient 
features of the time-scale dynamics of the detected coherent structures. The most interesting one is the existence of a
cascade process from larger scale towards smaller one, this process being continuous in time. This latter observation
indicates that coherent vorticity  structures play a crucial role in the time reversal symmetry breaking, a main feature
of the transition to turbulence.\\
Finaly, we wish to underline here, the similarities between our experimental results and
analytical models of coherent structures in the form of spiral vortices recently introduced by T.S. Lundgren
\cite{Lundgren}.  We would also like to underline the similarity between our experimental approach, and recent numerical
\cite{Farge98} and theoretical \cite{Arneo98} approaches relying on wavelet decompositions of the turbulent velocity field
(one-point measurement).

\newpage

\vspace{10mm}
{\bf ACKNOWLEDGMENTS}

\vspace{5mm}
This work is supported by the R\'egion Rh\^one-Alpes (Emergence project No. 97027229) and by Pr. J.P. Hansen, member of the Institut Universitaire de France. The authors would like to thank P. Flandrin, ENS-LYON, for stimulating discussions.


\newpage

\begin{center}
{\bf FIGURE CAPTIONS}
\end{center}

\vspace{5mm}

\begin{figure}
\centerline{
\epsfxsize=15cm
\epsffile{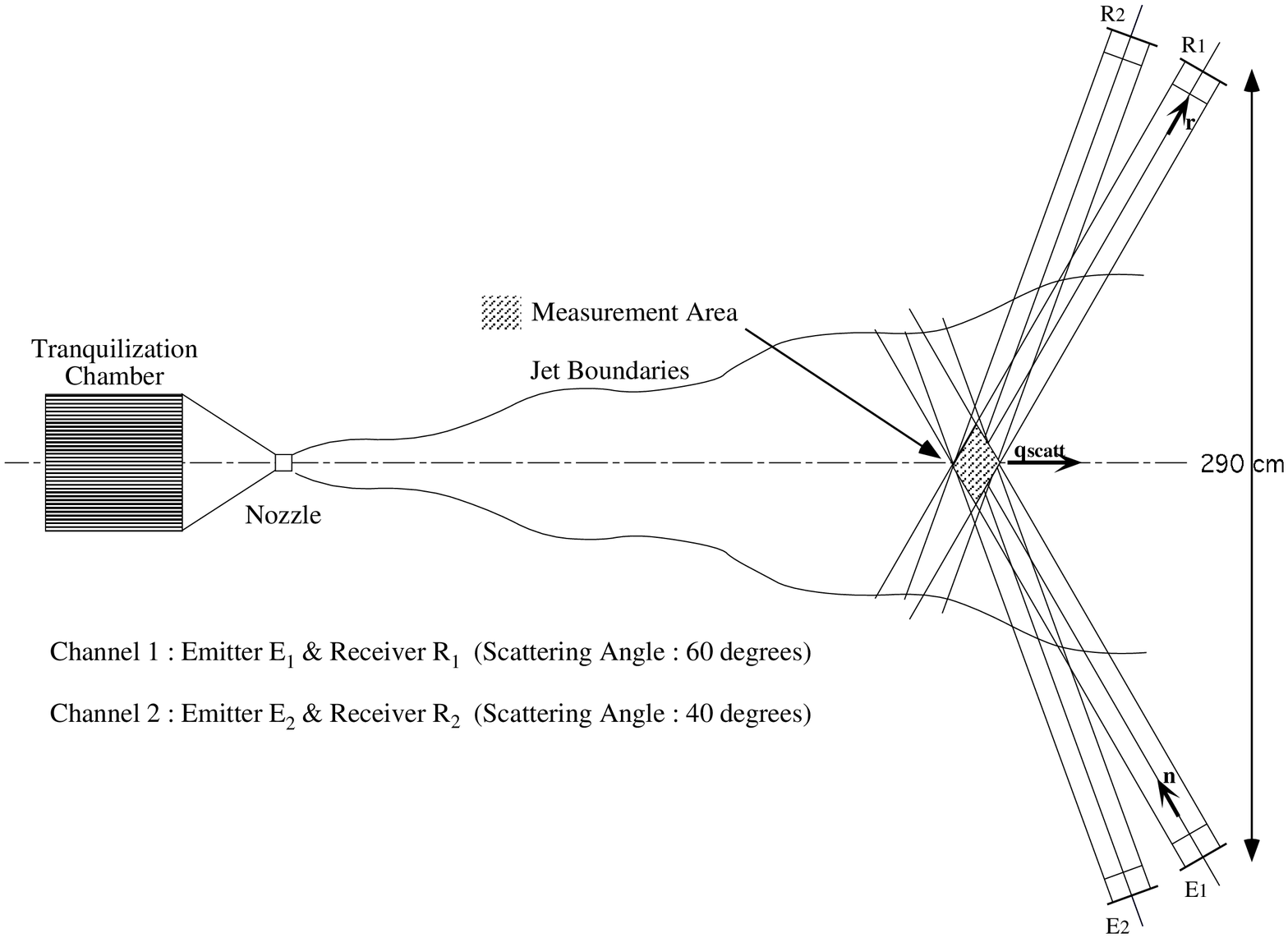}
}
\caption{ }
\label{setup}
\end{figure}

Figure \ref{setup}: Experimental setup.

\vspace{5mm}

\begin{figure}
\epsfxsize=15cm
\centerline{
\epsffile{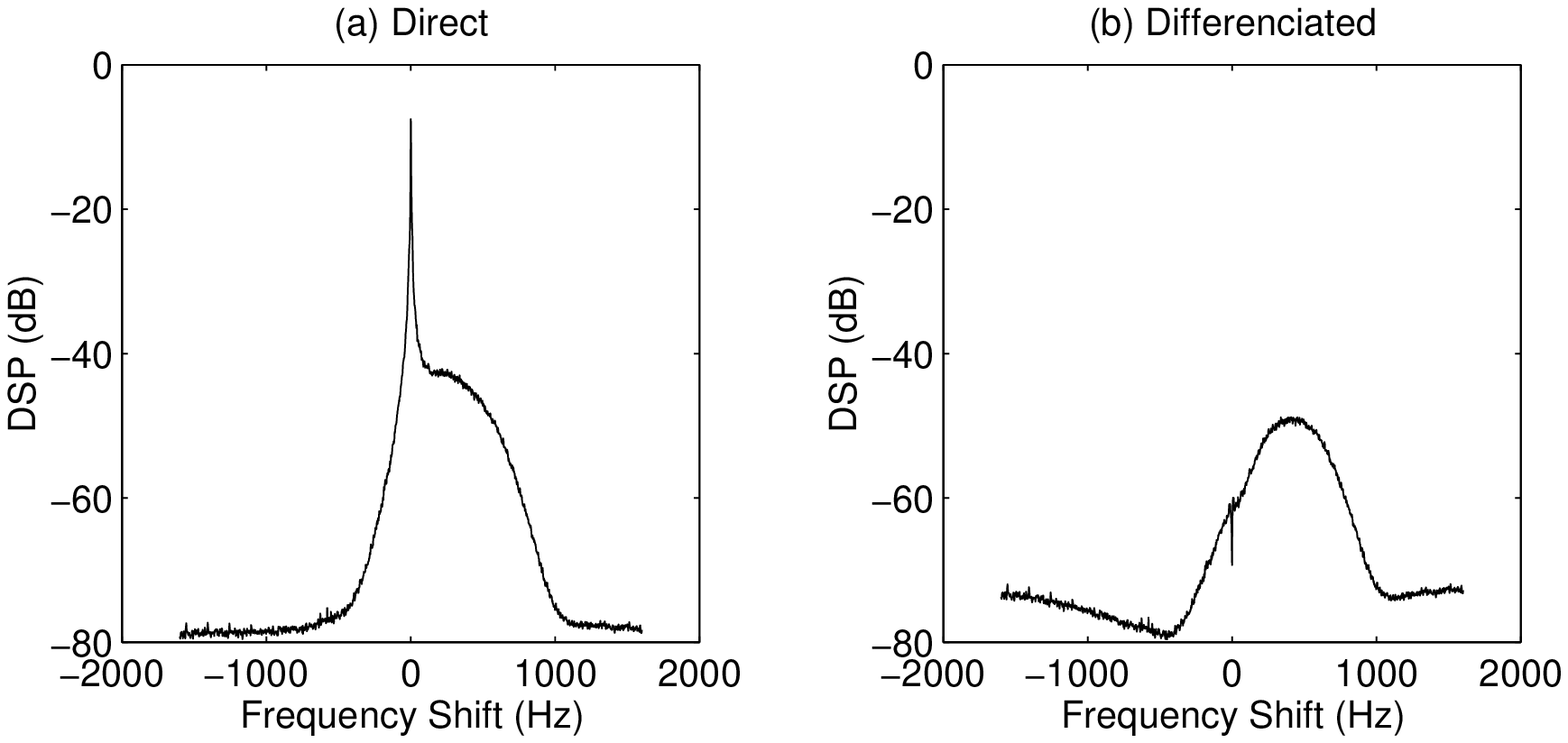}
}
\caption{ }
\label{spec_fig}
\end{figure}

Figure \ref{spec_fig}: Averaged PDS of direct and differentiated recorded data on  a jet-flow experiment. The
overall length of the data field was $2^{18}$ points . The PDS  is estimated from 2048 points data
segments with 1024 data points overlap.

\vspace{5mm}

\begin{figure}
\epsfxsize=15cm
\centerline{
\epsffile{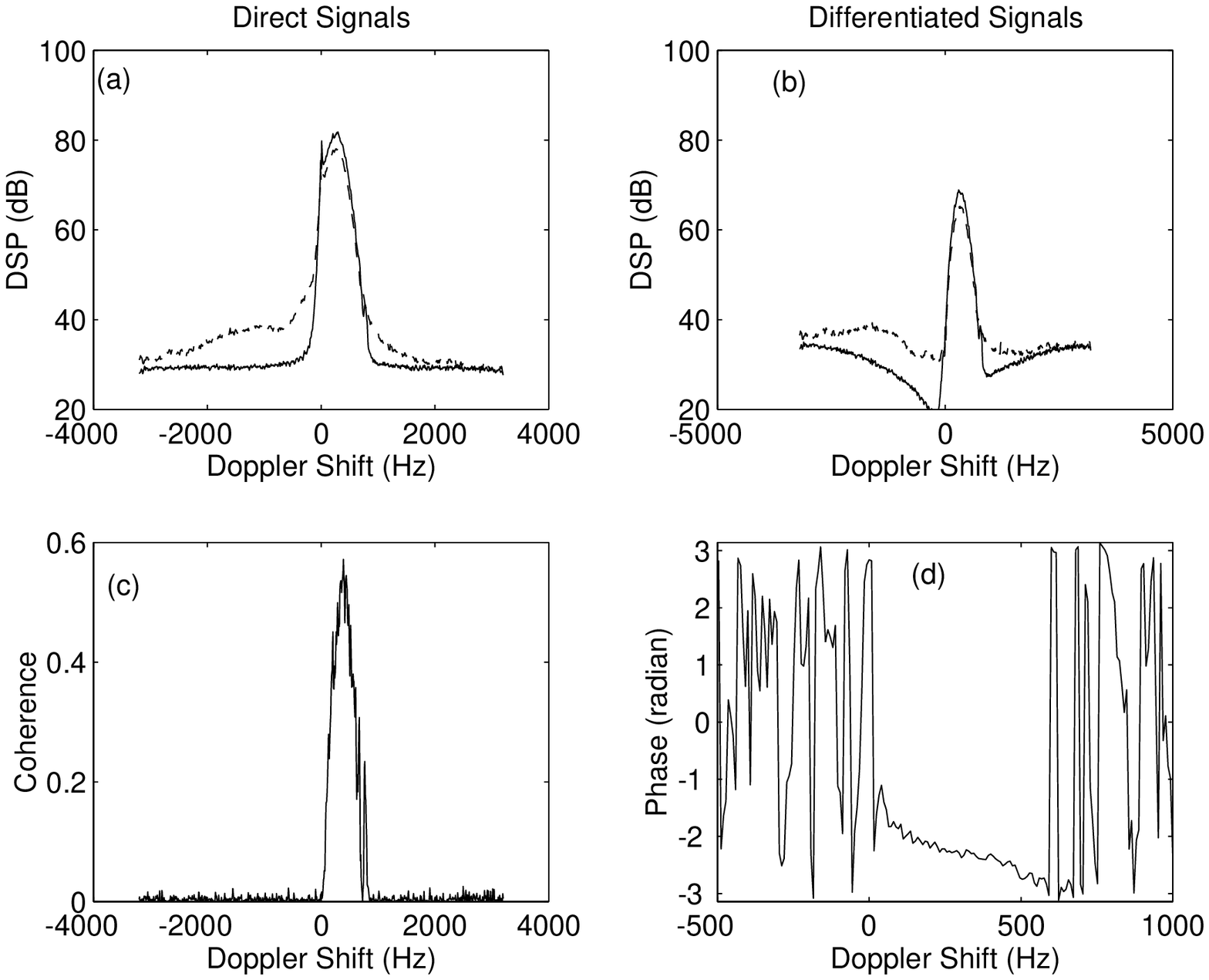}
}
\caption{ }
\label{figCoh}
\end{figure}

Figure \ref{figCoh}: (a) Estimated spectra from channels 1 and 2; (b) Estimated spectra of
differentiated signals recorded on channels 1 and 2; (c) Estimated cross-coherence function
between channels 1 and 2; (d) phase of the cross-channel interspectrum (notice that the frequency
span is smaller for this last plot).

\vspace{5mm}

\begin{figure}
\centerline{
\epsfxsize=15cm
\epsffile{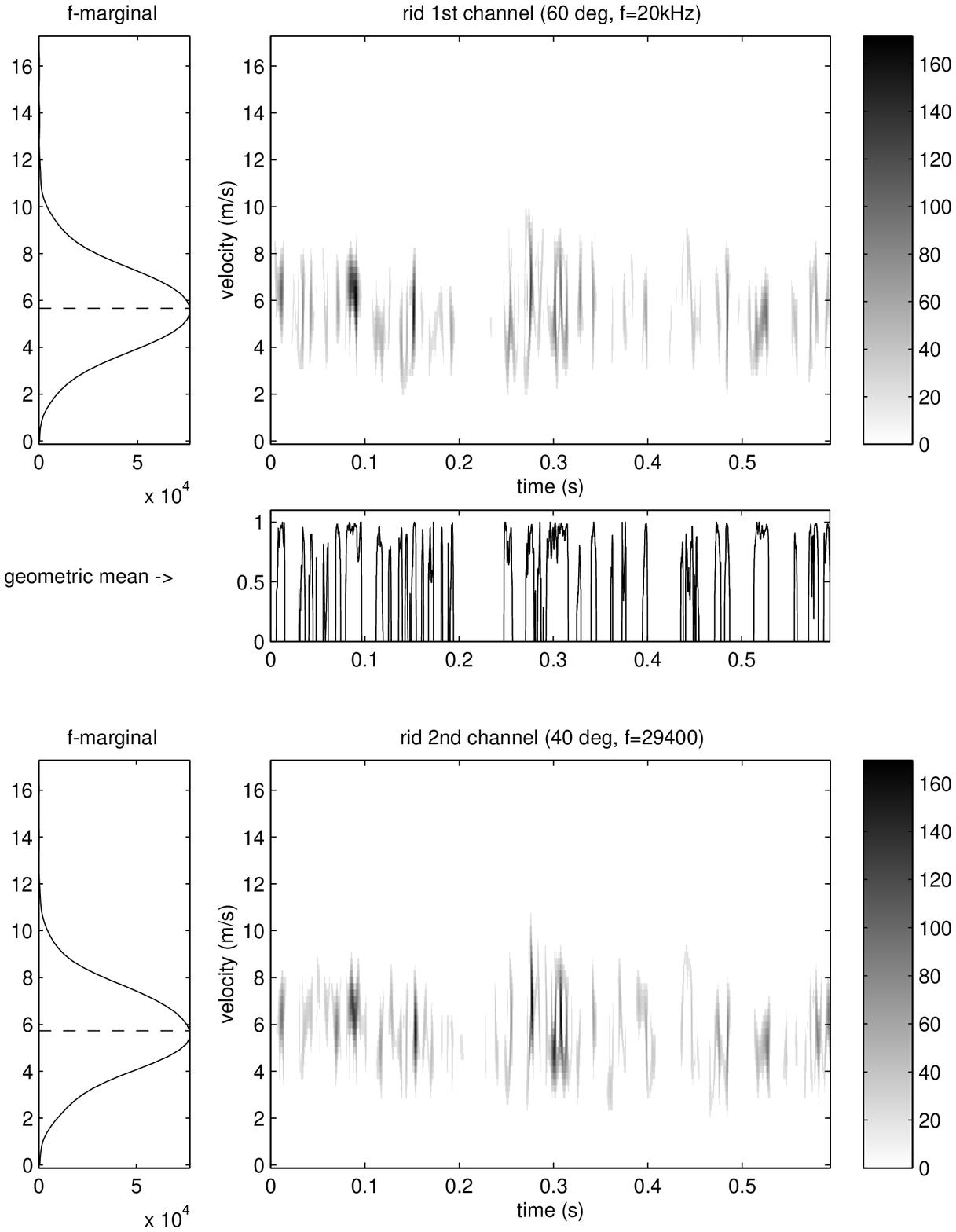}
}
\caption{ }
\label{fig20294}
\end{figure}

Figure \ref{fig20294}: 2 channels tf-rid estimation for approximate tuning conditions
(only positive values are represented).

\vspace{5mm}

\begin{figure}
\centerline{
\epsfxsize=10cm
\epsffile{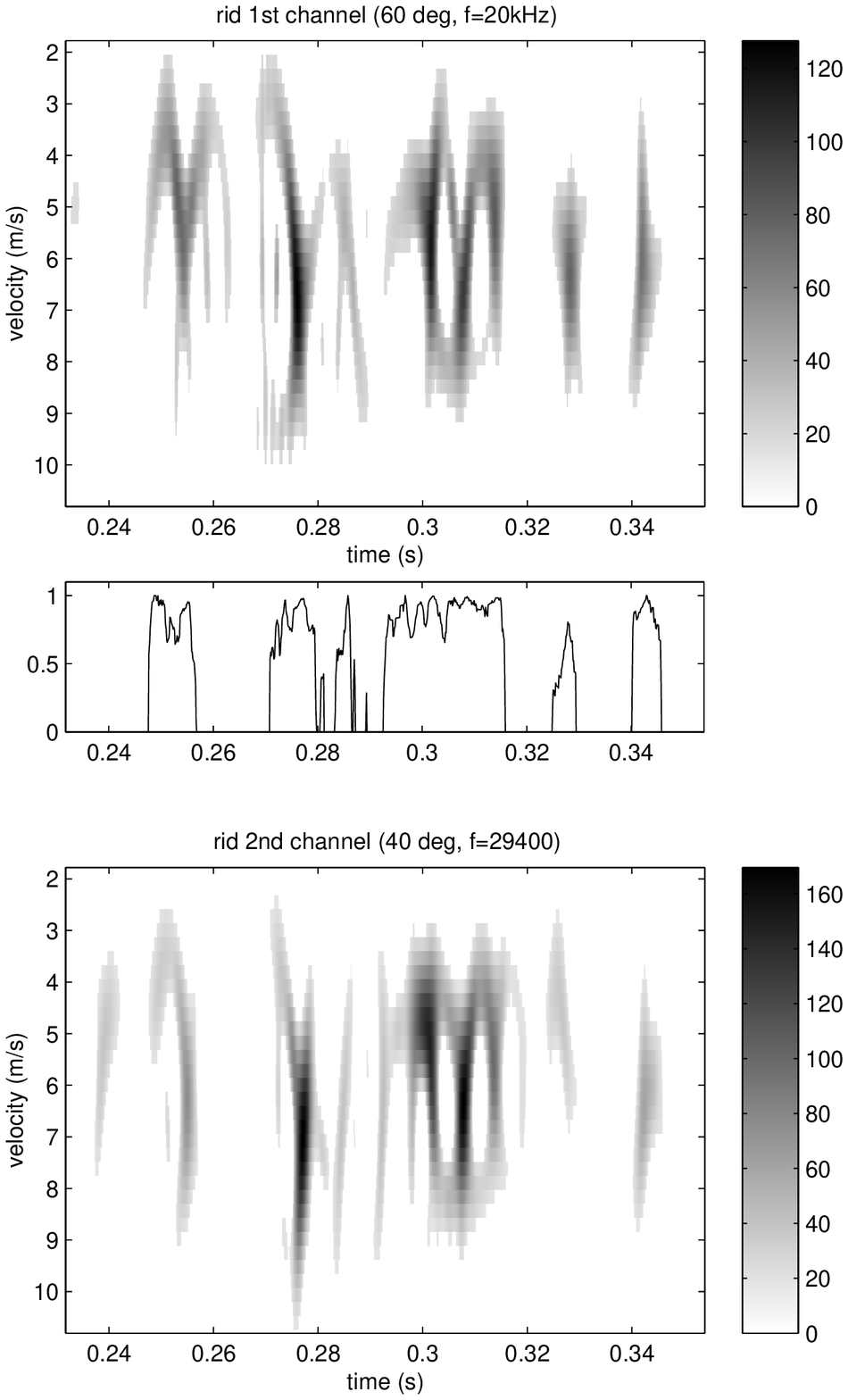}
}
\caption{ }
\label{fig20294zoom}
\end{figure}

Figure \ref{fig20294zoom}: 2 channels tf-rid estimation for approximate tuning conditions : details.

\vspace{5mm}

\begin{figure}
\centerline{
\epsfxsize=15cm
\epsffile{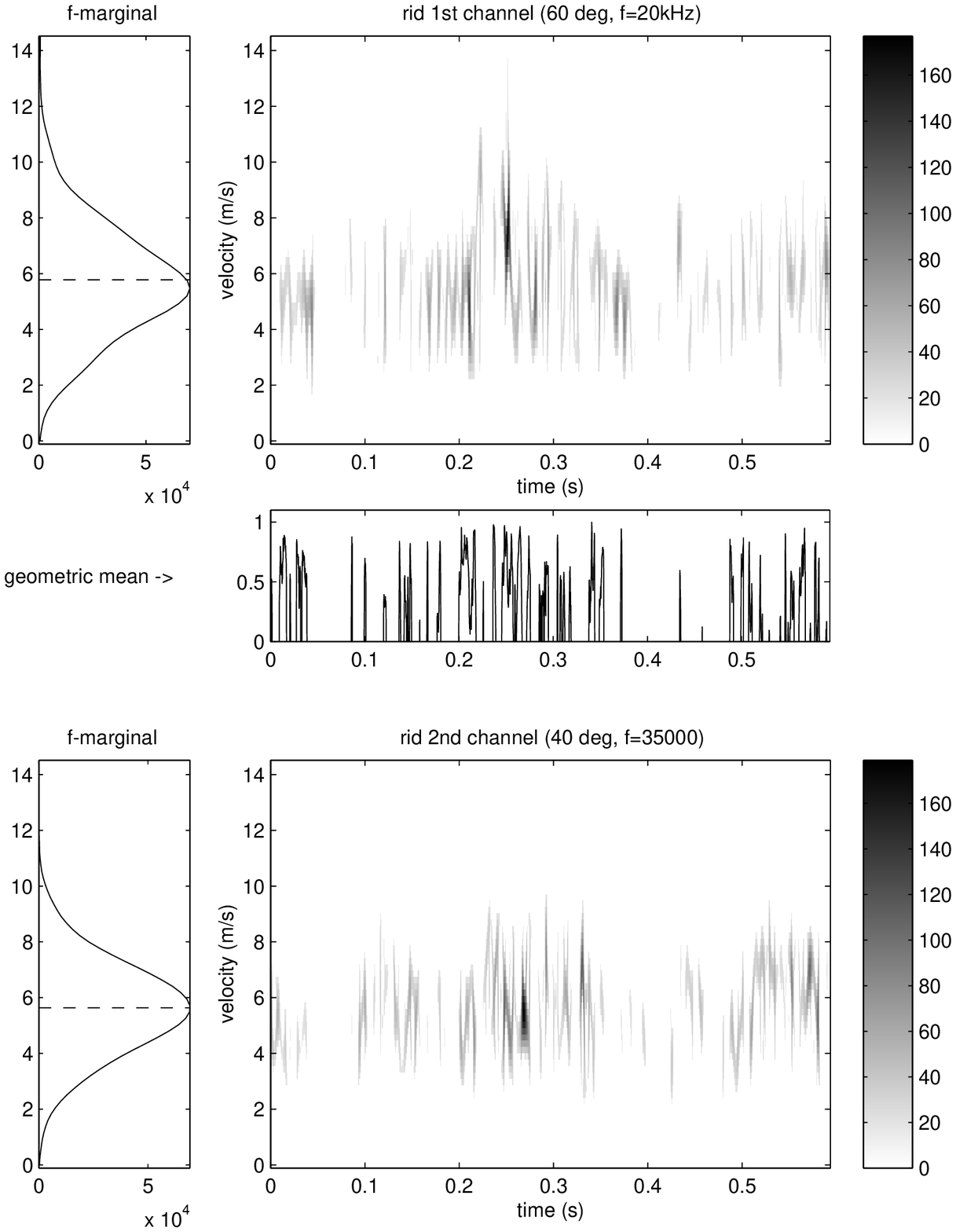}
}
\caption{ }
\label{fig20350}
\end{figure}

Figure \ref{fig20350}: 2 channels tf-rid estimation for $\vec{q}_{scat,1}=369,6 m^{-1}$ and  
$\vec{q}_{scat,2}=442,4 m^{-1}$
(only positive values are represented).

\vspace{5mm}

\begin{figure}
\centerline{
\epsfxsize=15cm
\epsffile{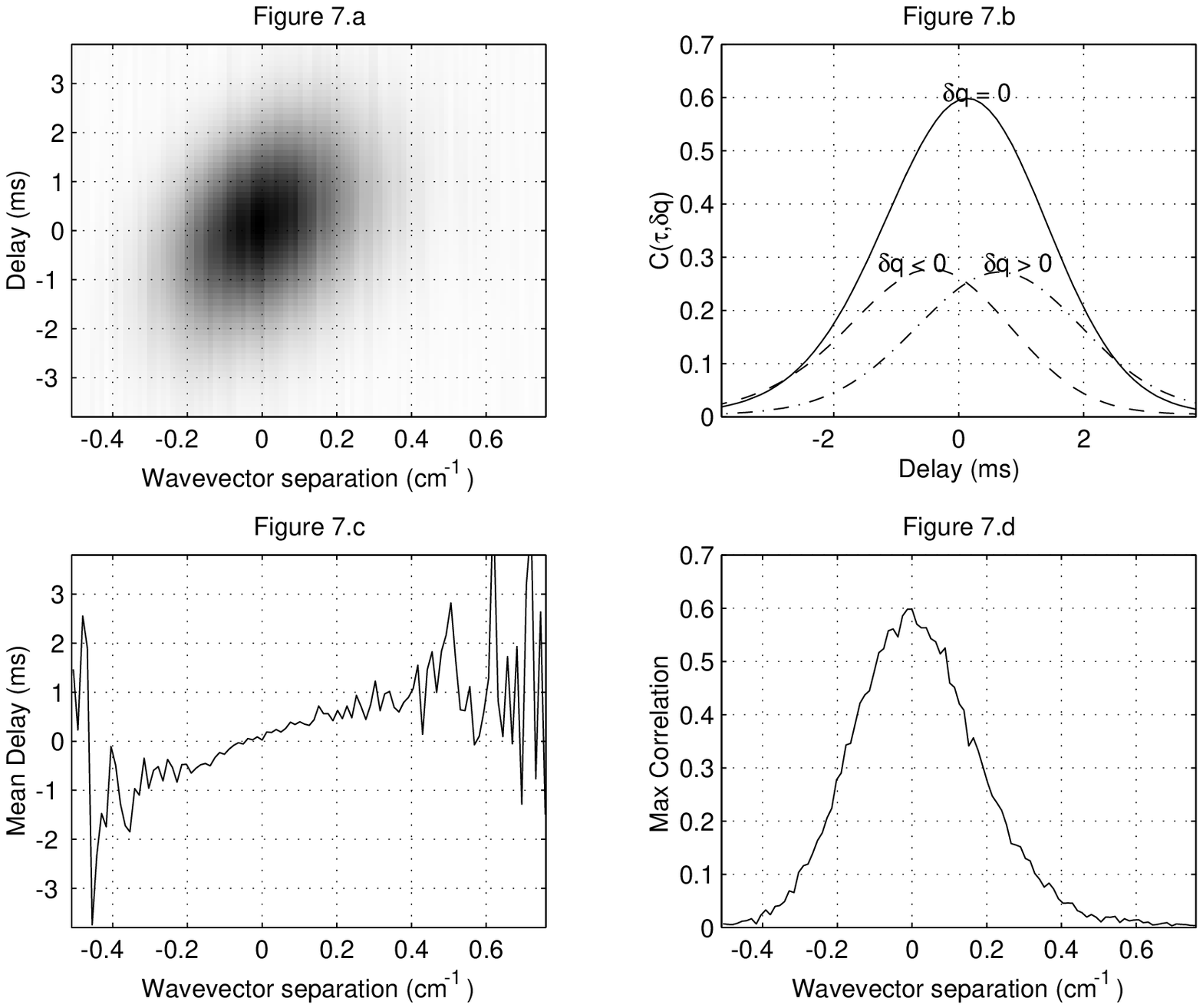}
}
\caption{ }
\label{fig_Corr}
\end{figure}

Figure \ref{fig_Corr}: Evolution of the time-scale correlation function $C(\tau,\delta q)$. (a) 2D plot
versus time delay $\tau$ and spatial wavectors separation $\delta q$; (b) Representations of $C(\tau,\delta q)$ 
for three different values of $\delta q$ ($0.0cm^{-1}, 0.2cm^{-1}, -0.2cm^{-1}$); (c) Plot of the mean delay time 
versus $\delta q$; (d) Plot of the maximum of $C(\tau,\delta q)$ versus $\delta q$.


\newpage

\newpage

\newpage

\newpage

\newpage 

\newpage

\newpage

\newpage

\begin{thebibliography}{99}
\bibitem{Frisch} U.Frisch, {\it Turbulence}, (Cambridge University Press, 1995).

\bibitem{Benzi93} R. Benzi, S. Ciliberto, C. Baudet \& G. Ruiz Chavarria, `` On the
scaling of three dimensional homogeneous and isotropic turbulence '', {\it Physica} D
{\bf80}, pp 385-398 (1993).

\bibitem{DCB91} S.Douady, Y.Couder, M.E.Brachet, `` Direct Observations of the
intermittency of intense vorticity filaments in turbulence '' Phys. Rev. Let. {\bf 67},
pp 983-986 (1991).

\bibitem{DPF96} B.Dernoncourt, J.F.Pinton, S.Fauve, `` Scaling of vorticity filaments in a turbulent swirling flow ''  Proceedings of the 6th European turbulence conference, Lausanne, pp 437-440 (1996).

\bibitem{Brac91} M.E.Brachet, `` Direct simulation of three-dimensional turbulence in
the Taylor-Green vortex '' Fluid. Dyn. Res. {\bf 8}, pp 1-8 (1991).

\bibitem{Les97} M.Lesieur. {\it Turbulence in fluids}, 3rd edition, Kluwer, Dordrecht
(1997).

\bibitem{Lev94}  Z.S. She \& E. L\'ev\^eque, `` Universal scaling laws in fully
developed turbulence '', {\it Phys. Rev. Lett.} {\bf72}, pp 336-339 (1994).

\bibitem{Obuk53} A.M. Obukhov, `` Effect of weak inhomogeneities in the atmosphere on
sound and light propagation '', {\it(Izv. Akad. Nauk. Seriya Geofiz.} {\bf2} pp 155-165
(1953).

\bibitem{Kraic53} R.H.Kraichnan, `` The scattering of Sound in a Turbulent Medium ''
,{\it J.Acoust. Soc. Am. } {\bf25} pp 1096-1104 (1983).
 
\bibitem{Chu58} B.T.Chu, L.S.G.Kov\`asznay, `` Non-linear interactions in a viscous
heat-conducting compressible gas '', {\it J. Fluid. Mech.} {\bf3} pp 494-514 (1958).

\bibitem{Lund} F. Lund \& C. Rojas, `` Ultrasound as a probe of turbulence '',{\it
Physica} D {\bf37}, pp 508-514 (1989).

\bibitem{will92} W.J.Williams, J.Jeong, in  {\it Time-frequency signal analysis: methods and applications}, Ed.~B.Bouashash, Longman \& Cheshire, Chapter~3, 74-97 (1992).

\bibitem{Batch53} G.K.Batchelor, {\it The theory of homogeneous turbulence}, Ed. Cambridge University Press,
Cambridge (1953).

\bibitem{Nelk94} M.Nelkin, ``Universality and scaling in fully developed turbulence'', 
Advances in  Physics, {\bf 43}, pp 143-181 (1994).

\bibitem{Noullez97} A.Noullez, G.Wallace, W.Lempert, R.B.Miles \& U.Frisch, ``Transverse velocity
increments in turbulent flows using the RELIEF technique'', {\it J. Fluid. Mech.}, {\bf338},
pp 287-307 (1997).

\bibitem{Wallace} J.M.Wallace, `` Methods for measuring vorticity in turbulent
flows '',{\it Experiments in Fluids}, {\bf 4}, pp 61-71 (1986).

\bibitem{Tsin92} A. Tsinober, E. Kit \& T. Dracos, `` Experimental investigation of the
field of velocity gradients in turbulent flows '', {\it J. Fluid. Mech.}, {\bf242},
pp 169-192 (1992).

\bibitem{Anton97} H. S. Shafi \& R. A. Antonia, `` Small-scale characteristics of a
turbulent boundary layer over a rough wall '', {\it J. Fluid. Mech.} {\bf342},
pp 263-293 (1997).

\bibitem{Colon} T.Colonius, S.K.Lele \& P.Moin, '' The scattering of sound waves by a
vortex: numerical simulations and analytical solutions ''{\it J. Fluid. Mech.}, Vol.260,
pp.271-298 (1994).

\bibitem{Engler} R. Engler et al., {\it J.Acoust. Soc. Am.} {\bf71} (1), pp 42-50
(1982), and {\it J.Acoust. Soc. Am. } {\bf85} (1) pp.72-82 (1989).

\bibitem{Kov76} C. M. Ho \& L.S.G. Kov\`asznay, `` Propagation of a coherent acoustic
wave through a turbulent shear flow '' ,{\it J.Acoust. Soc. Am. } {\bf60} pp
40-45 (1976).

\bibitem{Korm80} M.S. Korman \& R. T. Beyer, `` The scattering of sound by turbulence
in water '' ,{\it J.Acoust. Soc. Am. } {\bf67} (6) pp 1980-1987 (1976).

\bibitem{Baerg65} W. Baerg \& W.H. Schwarz, `` Measurements of the scattering of sound
from turbulence '' ,{\it J.Acoust. Soc. Am. } {\bf39} (6) pp 1125-1132 (1965).


\bibitem{Baud91} C.Baudet, S.Ciliberto \& J.F.Pinton,  `` Spectral analysis of the von
K\'arm\'an flow using ultrasound scattering '', Phys.Rev.Lett. {\bf 67-2}, pp 193-195
(1991).

\bibitem{Gromov82} P.R. Gromov, A.B.Ezerskii \& A.L.Fabrikant, `` Sound scattering by a vortex wake behind
a cylinder '', Sov.Phys.Acoust. {\bf 28-6}, pp 452-455 (1982).

\bibitem{BH96} C.Baudet, R.Hernandez, `` Spatial enstrophy spectrum in a fully
turbulent jet '', Proceedings of the 6th European turbulence conference,  Lausanne, pp
421-424, (1996).

\bibitem{MY75} A.S.Monin, A.M.Yaglom, {\it Statistical fluid mechanics}, (MIT Press, 1987).

\bibitem{Tennekes} H.Tennekes, J.L.Lumley, {\it A first course in turbulence}, (MIT Press, 1972).

\bibitem{Anke} D.Anke, Acustica {\bf 30}, (1974).

\bibitem{KFCS82} L.E.Kinsler, A.R.Frey, A.B.Coppens, J.V.Sanders, {\it Fundamentals of
acoustics}, (Wiley and Sons, 1982), 3rd edition.

\bibitem{Morse} P.M.Morse, K.U.Ingard,{\it Theoretical acoustics}, (Princeton University
Press, 1986).

\bibitem{wat66} W.A.Watkins,  Marine Bio-acoustics, {\bf 2}, 15-43 (1966).

\bibitem{choi89} H.I.Choi, W.J.Williams, IEEE Trans. Acoust., Speech, Signal Proc., {\bf 37}, No.~6, 862-871 (1989).

\bibitem{coh89} L.Cohen, Proc. IEEE, {\bf 77}, No.~7, 941-981 (1989).

\bibitem{cohen95} L. Cohen, {\it Time-frequency signal analysis,} Prentice Hall, 1995.

\bibitem{fla84} P.Flandrin, IEEE Int. Conf. Acoust., Speech, Signal Proc., {\bf 41B},
4.1-4.4, (1984).
 
\bibitem{Fla93} P.Flandrin, {\it Temps-fr\'equence}, Ed. Herm\`es, Paris (1993).

\bibitem{hla84} F.Hlawatsch,{\it  Digital Signal Processing - 84}, ed.~ V.Cappellini and A.Constantinides, 263-367, North-Holland (1984).

\bibitem{Jeong92} J. Jeong, W. J. Williams  IEEE Trans. Sig. Proc., {\bf 40}, No.~11, 2757-2765(1992).

\bibitem{hla92} F.Hlawatsh, G.F.Boudreaux-Bartels, IEEE Signal Processing Magazine,
{\bf 9}, No.~2, 21-67, (1992).

\bibitem{Arneo98} A. Arneodo, E. Bacry, S. Manneville \& J.F. Musy, `` Analysis of
random cascades using space-scale correlation functions '', {\it Phys. Rev. Lett}
{\bf80}, pp 708-711 (1998).

\bibitem{Farge98} M. Farge, K. Schneider \& N. K. R. Kevlahan,  `` Coherent structure
eduction in wavelet-forced two-dimensional turbulent flows '', in {\it Dynamics of slender vortices} Ed. E. Krause, Cambridge University Press (1998).

\bibitem{Lundgren} T.S. Lundgren, `` Strained spiral vortex model for turbulent fine structure '',
Phys. Fluids. {\bf 25}(12), pp 2193-2203 (1982). T.S. Lundgren, `` A small-scale turbulence model '',
Phys. Fluids. A {\bf 5}(6), pp 1472-1483 (1993). A.D. Gilbert, `` A cascade interpretation of Lundgren's streched spiral
vortex model for turbulent fine structure '', Phys. Fluids. A {\bf 5}(11), pp 2831-2834 (1993).

\end{thebibliography}
\end{document}